\documentstyle[axodraw,epsf]{elsart}
\begin{document}
\begin{frontmatter}
\title{
The Artificial Neural Networks as a tool for analysis of the
individual Extensive Air Showers data.
%Force line breaks with \\
}
\author{Tadeusz Wibig}
\address{Experimental Physics Dept., University of \L \'{o}d\'{z}, \\
ul. Pomorska 149/153, PL-90-236 \L \'{o}d\'{z}, Poland}

%\date{\today}

%\draft

%\maketitle

\begin{abstract}
In that paper we discuss possibilities of using the Artificial Neural Network
technic for the individual Extensive Air Showers data evaluation.
It is shown that the recently
developed new computational methods can be used in studies of EAS registered
by very large and complex detector systems.
The ANN can be used to classify
showers due to e.g. primary particle mass as well as to find a particular
EAS parameter like e.g. total muon number. The examples of both kinds of
analysis are given and discussed.

\end{abstract}

%\pacs{
%24.60.Ky, 13.85.Hd, 12.40.-y
%}
\end{frontmatter}

\section {Introduction}
The using of Artificial Neural Network (ANN) for solving a very different
physical problems become very extensive and promising in the last years
(\cite{ann}).
The stage of complexity of physical processes together with the
huge number of data to be proceed makes the ''classical type'' calculation
so complicated that the results sometimes can be reached on the very edge of
reality. One of the best examples of such situation is the Extensive Air
Shower (EAS) physics.
It is possible, in principle, to describe the EAS using the Monte--Carlo
technics. Main processes concerning particle traverse through the
air are known to some extend. Some believe that the contemporary knowledge
of high energy hadronic collisions is good enough to lead us to solutions
of some important cosmic ray physics problems. Even if it is so, the problem
is how to evaluate the physical answer from the cosmic ray experiment data.
The only one way possible at present is to compare the results of
Monte--Carlo calculations with the measurements. The most common
result of such comparison is the complain that the data obtained
from the measurement are not in a very close agreement with the
assumed model Monte--Carlo predictions. (Some ''physical''
discussions are about the question: ''How far they disagree?''. But this is not a
physical question!) The main goal is in the complexity of the problem. (The
Monte--Carlo programs used are about hundreds of thousands line long. The
first question is how much we can trust them. This is also not a
''physical'' question, but we should bear it in minds.) The best,
most popular I should say, Monte--Carlo codes to simulate a high
energy hadronic interactions have to contain two main sources of
uncertainty. First, due to the absence of the theory of strong
interactions to build models some more or less strong physical assumptions
about the interaction pictures are needed. Some simplifications
during the modeling is a natural way to do so and it can also be treated
as a source of uncertainty (''unremovable''). The another one is the values of
the model parameters which can not be taken from the theory and have to be
fitted to some (mainly accelerator) experiment results. The contemporary
Monte--Carlo
programs needs about hundred parameter (some of them very well known, some
unknown at all). The question arises: how the cosmic ray physics can be driven
from such a unpure ''theoretical'' predictions?
Let us look at that problem from the other side. The very best existing
(and those which will be build in near future) experimental apparatus
consist of large number of different detectors distributed over a wide
(effective) area. The data collected give a possibility to study
different characteristics of the EAS. Each of them is {\bf somehow} connected
to the different ''part'' of the shower development. The bold letters
were used to stress the connection which exists for sure, and which is
unknown for the reasons given above.
The question of the major importance in cosmic ray physics is about the
nature of
primary cosmic rays: energetic and mass spectrum. The detail knowledge of
the connections between cosmic ray flux on the top of the atmosphere and the
detector
response in the array on the ground level is certainly appreciated, but the
lack of it do not make a further study hopeless. Some general features
observed in EAS on ground level related to the mass of primary CR particle are
more or less {\em model independent\/}. Calculations shows that many of the
shower parameters depends on the mass but all those dependencies are smeared
due to individual shower development fluctuations. On the other hand, these
fluctuations are also related to the nature of the primary particle. To get
the maximum
information on the shower registered by a complex and extended array there are
at least two general ways.
First is to get a set of parameters describing the shower in the most
complete way. This can be understood as a contraction of the rough
experimental multidimensional space (in which each of the dimension is
given by a single detector signal) to the much less dimensional space
of some shower parameters. This reduction can be more or less fortunate
and there is no arbitrary way to do this. The further analysis of that
contracted space can be done in a conventional way by comparing the
experimental points with the Monte--Carlo simulated showers
(Ref.\ \cite{chili}).
In the case
of experiment dedicated to primary mass determination the most required
contraction is to reduce all measurement space to one--dimensional
which will be interpreted as a primary mass response of the apparatus.
Sometimes ones require that the reduction procedure should be also the
most effective one. This means that it should minimize the interpretation
errors. To find such best data evaluation procedure a different methods
can be used. The well defined one is the Principal Component Analysis
procedure discussed e.g. in Ref.\ \cite{pca}. The disadvantage of all that
procedures is in the
fact that all of them rely very strong on the Monte--Carlo simulation programs.
The proof of the exactness of the methods is very hard and is always connected
with the believe of the accuracy of the shower development description.

The second way is the object of that paper. The rough experimental data
space can be reduced promptly using the Artificial Neural Networks.
The ANN can be trained on a Monte--Carlo simulated array responses with the
known initial cosmic ray particle. The reduction of the dimensionality of
the output we wish to have is done by definition using
the general rules established in Monte--Carlo sample. The question of the
correctness
of the ANN procedure is more complicated mainly because the theory of such
a method is poorly known. The nature of the process of network self--organizing
is a enigma, and the rules developed by the network during training are
far from the standard ''physical'' ones. For the very complicated networks the
are even hard to extract. The proof of correctness is much harder than for a
known statistical methods, even if possible in general. But, as someone says,
''the proof of the pudding is in the eating'' I want to show that the method
can be satisfactory used in some cases.

In the present work I want to show some preliminary results concerning the
ANN analysis using as an example the muon part of the array of the KASCADE
experiment in Karlsruhe (Ref.\ \cite{kascade}). The muon array consists of
192 detectors of
3.2 $\mathrm {m^2}$ each spread over 200 $\times$ 200 m surface. The energy
threshold of muon is 0.1 GeV.

I want study to what extend the information
from these counters
can be analyzed with the help of ANN technics.

\section{Using ANN for the determination of particular shower parameters}

\begin{figure}
%\vspace{6cm}
\centerline{\epsfbox{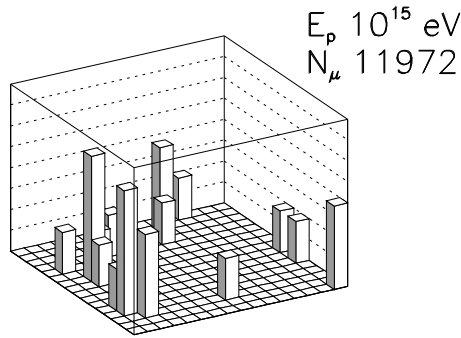}}
\caption{
The muon detector response for the typical shower initiated by primary
proton of energy of $\mathrm{10^15}$ eV in the array KASCADE experiment
geometry.
}
\label{easmi}
\end{figure}

The minimizing procedures usually used for
estimation of some distribution parameters need two conditions to be fulfilled.
One is to have a measurement of the shower statistically accurate enough
to be used. The problem can be seen
very well in Fig.\ \ref{easmi} where the typical muon detector responses for
the shower in the array of the KASCADE geometry is presented.

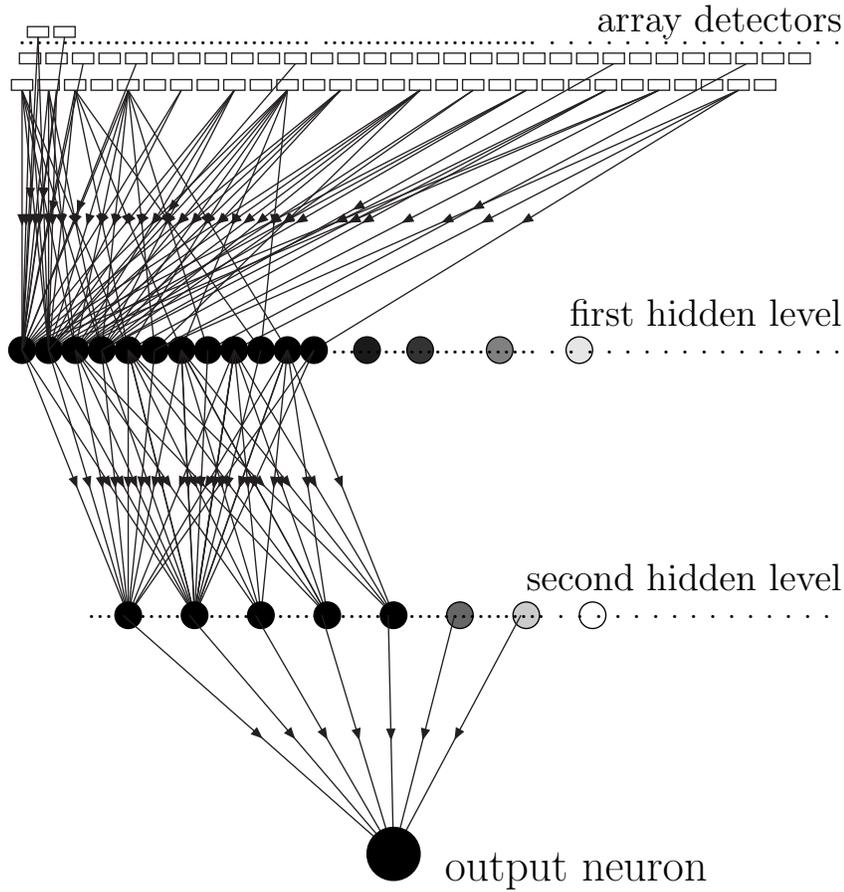
\begin{figure}
%%%%%%% ANN for Nm%%%%%%%%%%%%%%%%%
\begin{center}
\begin{picture}(320,420)(0,0)
\SetPFont{Helvetia}{6}
\SetScale{1.0}
\Boxc(10,300)(8,4)
\Boxc(20,300)(8,4)
\Boxc(30,300)(8,4)
\Boxc(40,300)(8,4)
\Boxc(50,300)(8,4)
\Boxc(60,300)(8,4)
\Boxc(70,300)(8,4)
\Boxc(80,300)(8,4)
\Boxc(90,300)(8,4)
\Boxc(100,300)(8,4)
\Boxc(110,300)(8,4)
\Boxc(120,300)(8,4)
\Boxc(130,300)(8,4)
\Boxc(140,300)(8,4)
\Boxc(150,300)(8,4)
\Boxc(160,300)(8,4)
\Boxc(170,300)(8,4)
\Boxc(180,300)(8,4)
\Boxc(190,300)(8,4)
\Boxc(200,300)(8,4)
\Boxc(210,300)(8,4)
\Boxc(220,300)(8,4)
\Boxc(230,300)(8,4)
\Boxc(240,300)(8,4)
\Boxc(250,300)(8,4)
\Boxc(260,300)(8,4)
\Boxc(270,300)(8,4)
\Boxc(280,300)(8,4)
\Boxc(290,300)(8,4)
\Boxc(13,310)(8,4)
\Boxc(23,310)(8,4)
\Boxc(33,310)(8,4)
\Boxc(43,310)(8,4)
\Boxc(53,310)(8,4)
\Boxc(63,310)(8,4)
\Boxc(73,310)(8,4)
\Boxc(83,310)(8,4)
\Boxc(93,310)(8,4)
\Boxc(103,310)(8,4)
\Boxc(113,310)(8,4)
\Boxc(123,310)(8,4)
\Boxc(133,310)(8,4)
\Boxc(143,310)(8,4)
\Boxc(153,310)(8,4)
\Boxc(163,310)(8,4)
\Boxc(173,310)(8,4)
\Boxc(183,310)(8,4)
\Boxc(193,310)(8,4)
\Boxc(203,310)(8,4)
\Boxc(213,310)(8,4)
\Boxc(223,310)(8,4)
\Boxc(233,310)(8,4)
\Boxc(243,310)(8,4)
\Boxc(253,310)(8,4)
\Boxc(263,310)(8,4)
\Boxc(273,310)(8,4)
\Boxc(283,310)(8,4)
\Boxc(293,310)(8,4)
\Boxc(303,310)(8,4)
\Boxc(16,320)(8,4)
\Boxc(26,320)(8,4)
\Text(320,317)[r]
{..................................
......................... . . . . . . . . . .   .   .     .
 .            .            .}
\Text(320,325)[r]{\large array detectors}
\ArrowLine(16,318)(10,200)
\ArrowLine(16,318)(20,200)
\ArrowLine(26,318)(10,200)
\ArrowLine(10,298)(10,200)
\ArrowLine(20,298)(10,200)
\ArrowLine(30,298)(10,200)
\ArrowLine(50,298)(10,200)
\ArrowLine(70,298)(10,200)
\ArrowLine(90,298)(10,200)
\ArrowLine(110,298)(10,200)
\ArrowLine(130,298)(10,200)
\ArrowLine(160,298)(10,200)
\ArrowLine(200,298)(10,200)
\ArrowLine(250,298)(10,200)
\GCirc(10,200){5}{0}
\ArrowLine(10,298)(20,200)
\ArrowLine(20,298)(20,200)
\ArrowLine(30,298)(20,200)
\ArrowLine(50,298)(20,200)
\ArrowLine(70,298)(20,200)
\ArrowLine(90,298)(20,200)
\ArrowLine(110,298)(20,200)
\ArrowLine(130,298)(20,200)
\ArrowLine(160,298)(20,200)
\ArrowLine(200,298)(20,200)
\ArrowLine(250,298)(20,200)
\GCirc(20,200){5}{0}
\ArrowLine(10,298)(30,200)
\ArrowLine(20,298)(30,200)
\ArrowLine(50,298)(30,200)
\ArrowLine(90,298)(30,200)
\ArrowLine(110,298)(30,200)
\ArrowLine(130,298)(30,200)
\ArrowLine(160,298)(30,200)
\ArrowLine(180,298)(30,200)
\ArrowLine(200,298)(30,200)
\ArrowLine(230,298)(30,200)
%\ArrowLine(250,298)(30,200)
\GCirc(30,200){5}{0}
\ArrowLine(10,298)(40,200)
\ArrowLine(20,298)(40,200)
\ArrowLine(50,298)(40,200)
\ArrowLine(90,298)(40,200)
\ArrowLine(110,298)(40,200)
\ArrowLine(130,298)(40,200)
\ArrowLine(160,298)(40,200)
\ArrowLine(180,298)(40,200)
\ArrowLine(230,298)(40,200)
%\ArrowLine(250,298)(40,200)
%\ArrowLine(280,298)(40,200)
\GCirc(40,200){5}{0}
\ArrowLine(10,298)(50,200)
\ArrowLine(30,298)(50,200)
\ArrowLine(50,298)(60,200)
\ArrowLine(110,298)(50,200)
\ArrowLine(160,298)(60,200)
\ArrowLine(230,298)(50,200)
\ArrowLine(280,298)(60,200)
\GCirc(50,200){5}{0}
\ArrowLine(30,298)(70,200)
\ArrowLine(50,298)(80,200)
\ArrowLine(110,298)(70,200)
\ArrowLine(280,298)(90,200)
\GCirc(60,200){5}{0}
\GCirc(70,200){5}{0}
\ArrowLine(33,308)(10,200)
\ArrowLine(53,308)(10,200)
\ArrowLine(113,308)(20,200)
\ArrowLine(233,308)(40,200)
\ArrowLine(283,308)(80,200)
\GCirc(80,200){5}{0}
\GCirc(90,200){5}{0}
\ArrowLine(30,298)(100,200)
\ArrowLine(50,298)(110,200)
\ArrowLine(110,298)(100,200)
\ArrowLine(250,298)(60,200)
\ArrowLine(280,298)(120,200)
\GCirc(100,200){5}{0}
\GCirc(110,200){5}{0}
\GCirc(120,200){5}{0}
\ArrowLine(16,298)(10,200)
\ArrowLine(13,298)(10,200)
\ArrowLine(13,298)(20,200)
\ArrowLine(23,298)(10,200)
\ArrowLine(23,298)(20,200)
\GCirc(140,200){5}{0.1}
\GCirc(160,200){5}{0.2}
\GCirc(190,200){5}{0.5}
\GCirc(220,200){5}{0.9}
\Text(320,200)[r]
{...........................
......................... . . . . . . . . . .   .   .     .
 .            .            .}
\Text(320,215)[r]{\large first hidden level}

\ArrowLine(10,200)(50,100)
\ArrowLine(20,200)(50,100)
\ArrowLine(30,200)(50,100)
\ArrowLine(40,200)(50,100)
\ArrowLine(50,200)(50,100)
\ArrowLine(60,200)(50,100)
\ArrowLine(70,200)(50,100)
\ArrowLine(90,200)(50,100)
\ArrowLine(110,200)(50,100)
\ArrowLine(120,200)(50,100)
\ArrowLine(10,200)(75,100)
\ArrowLine(20,200)(75,100)
\ArrowLine(30,200)(75,100)
\ArrowLine(40,200)(75,100)
\ArrowLine(50,200)(75,100)
\ArrowLine(70,200)(75,100)
\ArrowLine(80,200)(75,100)
\ArrowLine(90,200)(75,100)
\ArrowLine(100,200)(75,100)
\ArrowLine(110,200)(75,100)
\ArrowLine(120,200)(75,100)
\ArrowLine(90,200)(75,100)
\ArrowLine(50,200)(75,100)
\ArrowLine(70,200)(75,100)
\ArrowLine(90,200)(75,100)
\ArrowLine(110,200)(75,100)
\ArrowLine(90,200)(100,100)
\ArrowLine(110,200)(100,100)
\ArrowLine(30,200)(100,100)
\ArrowLine(70,200)(100,100)
\ArrowLine(50,200)(100,100)
\ArrowLine(70,200)(125,100)
\ArrowLine(30,200)(125,100)
\ArrowLine(50,200)(125,100)
\ArrowLine(70,200)(125,100)
\ArrowLine(90,200)(125,100)
\ArrowLine(110,200)(125,100)
\ArrowLine(50,200)(150,100)
\ArrowLine(70,200)(150,100)
\ArrowLine(90,200)(150,100)
\ArrowLine(110,200)(150,100)
\GCirc(50,100){5}{0}
\GCirc(75,100){5}{0}
\GCirc(100,100){5}{0}
\GCirc(125,100){5}{0}
\GCirc(150,100){5}{0}
\GCirc(175,100){5}{0.4}
\GCirc(200,100){5}{0.8}
\GCirc(225,100){5}{1}
\Text(320,100)[r]
{...........................
......................... . . . . . . . . . .   .   .     .
 .            .   }
\Text(320,115)[r]{\large second hidden level}

\ArrowLine(48,100)(150,010)
\ArrowLine(73,100)(150,010)
\ArrowLine(98,100)(150,010)
\ArrowLine(123,100)(150,010)
\ArrowLine(148,100)(150,010)
\ArrowLine(173,100)(150,010)
\ArrowLine(198,100)(150,010)
\GCirc(150,010){10}{0}
\Text(170,000)[lb]{\Large output neuron}
\end{picture}
\end{center}
%%%%%%%%%%%%%%%%%%%%%%%%%%%%%%%%%%%%%%%%
\caption{
Schematic layout of the Artificial Neural Network used for the total number
of muons in individual EAS evaluation
}
\label{annnm}
\end{figure}

In the most of showers the numbers
registered by the muon detectors oscillated around the very few. The large
statistical density fluctuations are expected.
Another
condition is that the assumption about the real distribution one trying
to fit is a correct one. In literature there are usually a few possibilities
and which one to chose is a question of taste. The differences is expected
not to be large, but a word ''large'' has not a precise meaning.
The ANN method is, by definition, not bothered by both such problems.
To show that,
I want to present the results concerning the total muon number determination
for individual showers by the KASCADE--like experiments.
It should be pointed out here also that the traditional minimization methods
of the muon lateral distribution determination need additional information
about the Extensive Air Shower usually obtained from the measurement of the
electromagnetic component part of the experiment. They are the shower core
position and the shower axis inclination angles. For the results presented
in that section those shower parameters are not used. The estimation
of the muon size of the shower using ANN is based only on the muon
component registration.
Then schematic view of the ANN architecture is given in Fig. \ref{annnm}.

\begin{figure}
\vspace{6cm}
\epsfxsize 250pt
\hspace{-1.5cm}
\centerline{
\epsfbox[0 0 600 150]
{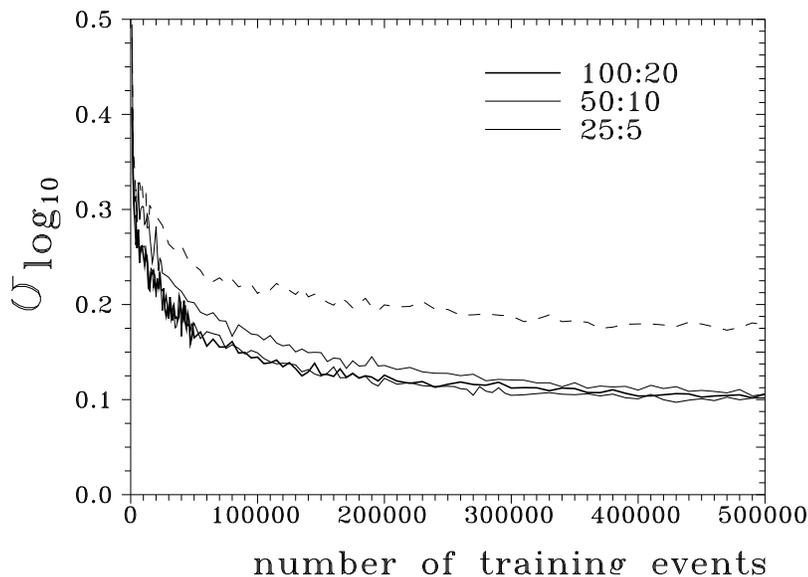}}
%\centerline{\epsfbox{annsignm.eps}}
\caption{
The efficiency of the ANN as a function of the number of events using for
the training process. Different solid lines shows the results for different
numbers of neurons in two hidden levels. The dashed line is a result for
incline showers
}
\label{nmt}
\end{figure}

The input contain 192 signals from the ideal
detectors measured the numbers of muons passing the detector surface.
Each of the inputs is connected with each
of the first hidden level neurons. The analysis was performed
the network with the two hidden levels with different number of neurons
to see the effect of the network size. The last hidden level is connected
to the one output unit. The number of the network parameters to be trained
was from about 25000 to 500000. As the response function the
common sigmoid function

\begin{equation}
\mathrm{output} \ = \ {{1} \over {1 + {\mathrm A\: {e}}^{- \mathrm{input}}}}
\label{sigmoid}
\end{equation}

was used and the training was the standard back--propagation method.
To reach the results about hundreds of thousands showers have been used
for training procedure.
The direct use of the Monte--Carlo simulation program for training was
simply impossible so the special pseudo--Monte--Carlo generator was
developed. The semi--empirical description of
CORSIKA v4.112 (Ref.\ \cite{corsika})
showers was used. From the point of view of total muon number estimation
the details are not very important. For the same reason only vertical
showers were used hereafter the present analysis. Results for incline
showers do not differ much. An example is given in Fig. \ref{nmt}.
It should be said that for the final ANN tests the exact CORSIKA output
showers were used. The results presented below justify thus the exactness
of our pseudo--Monte--Carlo shower generator.

In the Fig. \ref{nmt} the convergency of the training procedure is shown
for different network sizes. On the vertical axis the width of the
distribution of the deviation of the estimated total muon number from its
true value is given.

As it is seen the further enlargement of the network size do not lead to the
improvement of the accuracy of the ANN answers. It should be noted that the
ANN works for incline showers quite well as well. The fluctuations on the
single detectors for the incline showers increase due to the cos($\Theta$)
factor reduction of the detector effective area.

\begin{figure}
\epsfxsize 250pt
\hspace{-1.5cm}
\centerline{
\epsfbox[0 0 600 600]{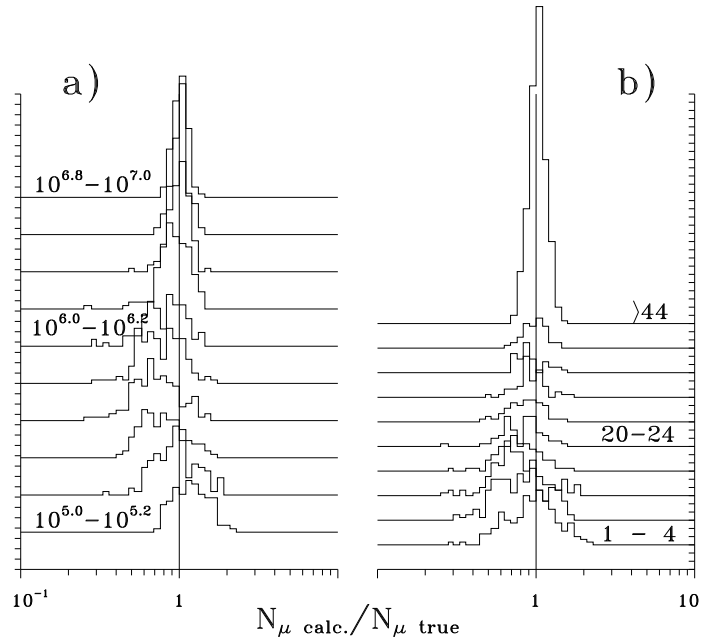}}
\caption{
The accuracy of determining the total muon number in individual showers
shown for different shower sizes labeled by the primary particle particle
energy a) and by the number of hit detectors in the KASCADE--like array b).
}
\label{nm1}
\end{figure}

\begin{figure}
\epsfxsize 250pt
\hspace{-1.5cm}
\centerline{
\epsfbox[0 0 600 600]{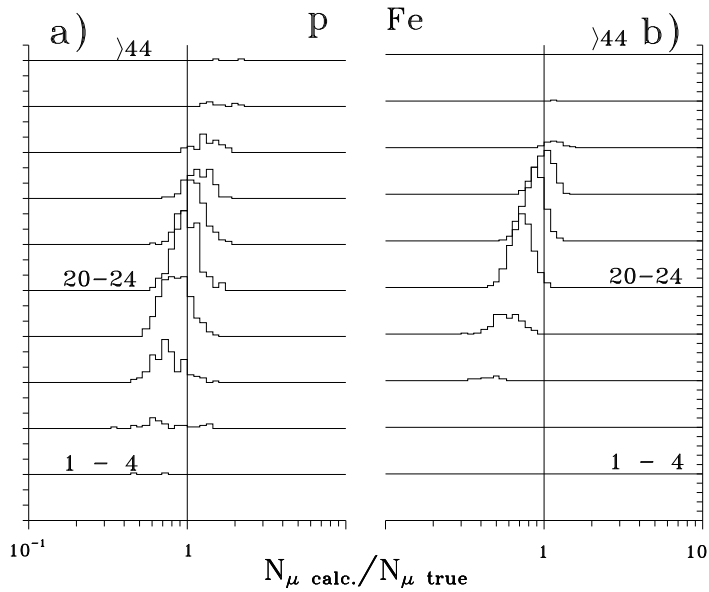}}
\caption{
The accuracy of determining the total muon number in individual showers
initiated by primary protons a) and iron nuclei b) of the energy of
$\mathrm 10^{15}$ eV. The histograms for different number of hit detectors are
presented separately as in Fig. 4b.
}
\label{nm6}
\end{figure}

\begin{figure}
\epsfxsize 250pt
\hspace{-1.5cm}
\centerline{
\epsfbox[0 0 600 600]{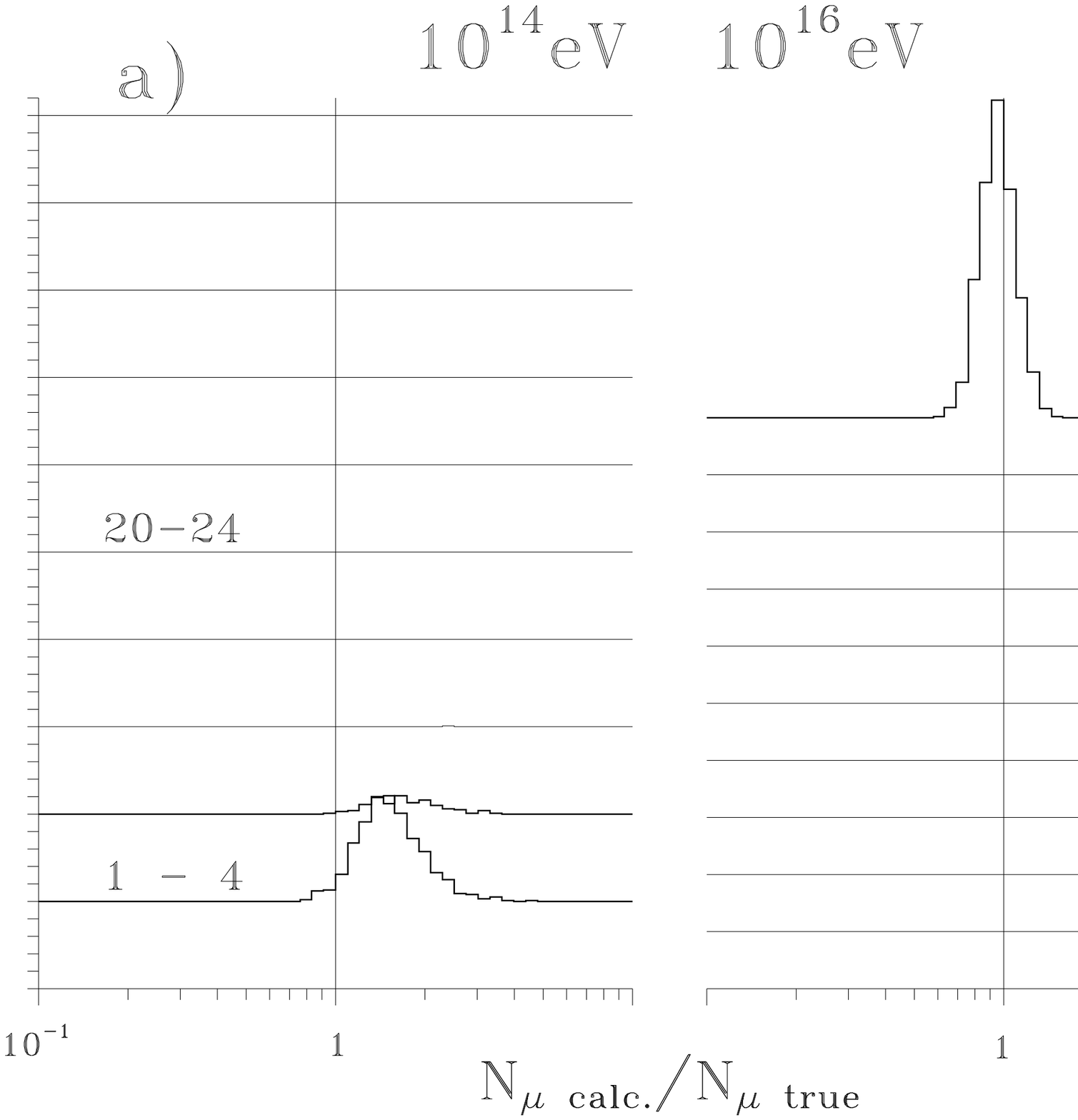}}
\caption{
The same as in Fig.\ 4
%\ \ref{nm}
for very small a) and relatively large b) proton showers
( of energies $10^{14}$ and $10^{16}$ eV respectively ).
}
\label{nm57}
\end{figure}

The obtained results are presented in Figs. \ref{nm1}--\ref{nm57}.

The accuracy seen in the muon size determination
is quite good. The interesting point is that the network trained with the
proton showers only gives in the tests some answers for the iron induced
showers as it is seen in Fig. \ref{nm6}. The bias toward the smaller
values is seen what is clearly the result of a difference in shapes of
proton and iron muon lateral distributions. That bias disappears when the
training procedure contains also the heavy primaries in the primary
particle spectrum.

Another important feature of the ANN method is presented in
Fig.\ \ref{nm57}. The ANN is able to give an answer also when only
very few detectors are hit. This is presented in Fig.\ \ref{nm57}a.

The bias seen in Fig.\ \ref{nm57}a is expected due to existence of the
threshold minimum 4 hit detectors in the training shower sample.

The comparison of the ANN results with the standard minimization technics
based method is given in Ref.\ \cite{minim}. There is shown that the
spread of the estimated total muon number with the respect to the true one
obtained using the ANN procedure is about the same as for the best
minimization methods for the relatively large showers when the statistical
weight of an information collected by the detectors is high enough and
is much smaller for the small and very small showers.

\section{Using ANN for the determination of primary particle mass}

It is clear that the primary cosmic ray particle mass is involved in
the air shower development. The possible mass spectrum consists of course
in principle all the stable nuclides (in practice from hydrogen to iron).
However the fluctuations of the cascading process makes impossible to
distinguish the very close masses and because of that the primary
cosmic ray spectrum is often studied like a spectrum of the group of
similar nuclei. The whole mass spectrum is divided into the five groups:
H, He, light (C-N-O), heavy (about Si) and very heavy (Fe). The relative
abundances of these groups is known as the question of the cosmic ray
primary mass spectrum.
For such task the network architecture was changed. Two different possibilities
have been examined. In the first the preproceed data were used as an inputs.
Instead of direct muon detector
responses a few parameter of the muon lateral distribution were used. The
number of 192 input nodes with an information about the muons was reduced to
only four: muon densities at two well measured distances (50 and 100 m) and the
relative slopes of the muon lateral distribution these distances. The
important information
about the primary mass composition is also included in the electromagnetic
shower component so another four input nodes were introduced with the same as
for muon information but about the electron lateral distribution. To increase
the number of intranetwork connections the additional hidden level was
introduced.
As an output level instead of one the five neurons were established
each one
related to the one group of nuclei in the primary cosmic ray mass spectrum.
As the ANN answer the number of the neuron of the highest output signal was
chosen.

\begin{figure}
%\vspace{6cm}
\begin{center}
\begin{picture}(320,420)(0,0)
\SetPFont{Helvetia}{12}
\SetScale{1.0}
\Boxc(30,340)(50,10)
\Boxc(80,341)(50,10)
\Boxc(130,342)(50,10)
\Text(0,360)[l]{electron detector information}
\Boxc(190,340)(50,10)
\Boxc(240,341)(50,10)
\Boxc(290,342)(50,10)
\Text(320,360)[r]{muon detector information}
\ArrowLine(30,335)(10,200)
\ArrowLine(80,336)(10,200)
\ArrowLine(130,337)(10,200)
\ArrowLine(30,335)(20,200)
\ArrowLine(80,336)(20,200)
\ArrowLine(130,337)(20,200)
\ArrowLine(30,335)(30,200)
\ArrowLine(80,336)(30,200)
\ArrowLine(130,337)(30,200)
\ArrowLine(30,335)(40,200)
\ArrowLine(80,336)(40,200)
\ArrowLine(130,337)(40,200)
\ArrowLine(30,335)(50,200)
\ArrowLine(80,336)(50,200)
\ArrowLine(130,337)(50,200)
\ArrowLine(30,335)(60,200)
\ArrowLine(80,336)(70,200)
\ArrowLine(130,337)(70,200)
\ArrowLine(30,335)(70,200)
\ArrowLine(80,336)(100,200)
\ArrowLine(130,337)(110,200)
\ArrowLine(190,335)(10,200)
%\ArrowLine(240,336)(10,200)
%\ArrowLine(290,337)(10,200)
\ArrowLine(190,335)(20,200)
\ArrowLine(240,336)(20,200)
\ArrowLine(290,337)(20,200)
\ArrowLine(190,335)(30,200)
\ArrowLine(240,336)(30,200)
%\ArrowLine(290,337)(30,200)
\ArrowLine(190,335)(40,200)
\ArrowLine(240,336)(40,200)
\ArrowLine(290,337)(40,200)
\ArrowLine(190,335)(50,200)
\ArrowLine(240,336)(50,200)
%\ArrowLine(290,337)(50,200)
\ArrowLine(190,335)(60,200)
\ArrowLine(240,336)(70,200)
\ArrowLine(290,337)(70,200)
\ArrowLine(190,335)(70,200)
\ArrowLine(240,336)(100,200)
\ArrowLine(290,337)(100,200)
\ArrowLine(290,337)(120,200)
\GCirc(10,200){5}{0}
\GCirc(20,200){5}{0}
\GCirc(30,200){5}{0}
\GCirc(40,200){5}{0}
\GCirc(50,200){5}{0}
\GCirc(60,200){5}{0}
\GCirc(70,200){5}{0}
\GCirc(80,200){5}{0}
\GCirc(90,200){5}{0}
\GCirc(100,200){5}{0}
\GCirc(110,200){5}{0}
\GCirc(120,200){5}{0}
\GCirc(140,200){5}{0.1}
\GCirc(160,200){5}{0.2}
\GCirc(190,200){5}{0.5}
\GCirc(220,200){5}{0.9}
\Text(320,200)[r]
{...........................
......................... . . . . . . . . . .   .   .     .
 .            .     }
\Text(320,215)[r]{\large first hidden level}

\ArrowLine(10,200)(25,144)
\ArrowLine(20,200)(25,144)
\ArrowLine(30,200)(25,144)
\ArrowLine(40,200)(25,144)
\ArrowLine(50,200)(25,144)
\ArrowLine(60,200)(25,144)
\ArrowLine(70,200)(25,144)
\ArrowLine(90,200)(25,144)
\ArrowLine(110,200)(25,144)
%\ArrowLine(140,200)(25,144)
\ArrowLine(10,200)(50,144)
\ArrowLine(20,200)(50,144)
\ArrowLine(30,200)(50,144)
\ArrowLine(40,200)(50,144)
\ArrowLine(50,200)(50,144)
\ArrowLine(60,200)(50,144)
\ArrowLine(70,200)(50,144)
\ArrowLine(90,200)(50,144)
\ArrowLine(110,200)(50,144)
%\ArrowLine(140,200)(50,144)
\ArrowLine(10,200)(75,144)
\ArrowLine(20,200)(75,144)
\ArrowLine(30,200)(75,144)
\ArrowLine(40,200)(75,144)
\ArrowLine(50,200)(75,144)
\ArrowLine(70,200)(75,144)
%\ArrowLine(140,200)(75,144)
\ArrowLine(90,200)(75,144)
\ArrowLine(110,200)(75,144)
\ArrowLine(10,200)(100,144)
\ArrowLine(30,200)(100,144)
\ArrowLine(70,200)(100,144)
\ArrowLine(50,200)(100,144)
\ArrowLine(70,200)(125,144)
\ArrowLine(50,200)(125,144)
\ArrowLine(90,200)(150,144)
\ArrowLine(110,200)(150,144)
\GCirc(25,144){5}{0}
\GCirc(50,144){5}{0}
\GCirc(75,144){5}{0}
\GCirc(100,144){5}{0}
\GCirc(125,144){5}{0}
\GCirc(150,144){5}{0}
\GCirc(175,144){5}{0.4}
\GCirc(200,144){5}{0.8}
\GCirc(225,144){5}{1}
\Text(320,144)[r]
{...........................
......................... . . . . . . . . . .   .   .     .
 .            .            .}
\Text(320,159)[r]{\large second hidden level}

\ArrowLine(25,144)(50,89)
\ArrowLine(50,144)(50,89)
\ArrowLine(75,144)(50,89)
\ArrowLine(100,144)(50,89)
\ArrowLine(125,144)(50,89)
\ArrowLine(150,144)(50,89)
%\ArrowLine(175,144)(50,89)
%\ArrowLine(200,144)(50,89)
\ArrowLine(25,144)(75,89)
\ArrowLine(50,144)(75,89)
\ArrowLine(75,144)(75,89)
\ArrowLine(100,144)(75,89)
\ArrowLine(125,144)(75,89)
\ArrowLine(150,144)(75,89)
%\ArrowLine(175,144)(75,89)
%\ArrowLine(200,144)(75,89)
\ArrowLine(25,144)(100,89)
\ArrowLine(50,144)(100,89)
\ArrowLine(100,144)(100,89)
%\ArrowLine(200,144)(100,89)
\ArrowLine(25,144)(125,89)
\ArrowLine(50,144)(125,89)
\ArrowLine(100,144)(125,89)
\ArrowLine(150,144)(125,89)
\ArrowLine(50,144)(150,89)
\ArrowLine(25,144)(150,89)
\ArrowLine(125,144)(150,89)
\ArrowLine(150,144)(150,89)
\GCirc(50,89){5}{0}
\GCirc(75,89){5}{0}
\GCirc(100,89){5}{0}
\GCirc(125,89){5}{0}
\GCirc(150,89){5}{0}
\GCirc(175,89){5}{0.4}
\GCirc(200,89){5}{0.8}
\GCirc(225,89){5}{1}
\Text(320,89)[r]
{...........................
......................... . . . . . . . . . .   .   .     .
 .  }
\Text(320,104)[r]{\large third hidden level}

\ArrowLine(50,89)(50,021)
\ArrowLine(75,89)(50,021)
\ArrowLine(100,89)(50,021)
\ArrowLine(125,89)(50,021)
\ArrowLine(150,89)(50,021)
%\ArrowLine(175,89)(50,021)
%\ArrowLine(200,89)(50,021)
\ArrowLine(50,89)(100,021)
\ArrowLine(75,89)(100,021)
\ArrowLine(100,89)(100,021)
\ArrowLine(125,89)(100,021)
\ArrowLine(150,89)(100,021)
%\ArrowLine(175,89)(100,021)
%\ArrowLine(200,89)(100,021)
\ArrowLine(50,89)(150,021)
\ArrowLine(75,89)(150,021)
\ArrowLine(100,89)(150,021)
\ArrowLine(125,89)(150,021)
\ArrowLine(150,89)(150,021)
%\ArrowLine(175,89)(150,021)
%\ArrowLine(200,89)(150,021)
\ArrowLine(50,89)(200,021)
\ArrowLine(75,89)(200,021)
\ArrowLine(100,89)(200,021)
\ArrowLine(150,89)(200,021)
\ArrowLine(125,89)(200,021)
%\ArrowLine(175,89)(200,021)
%\ArrowLine(200,89)(200,021)
\ArrowLine(50,89)(250,021)
\ArrowLine(75,89)(250,021)
\ArrowLine(100,89)(250,021)
\ArrowLine(125,89)(250,021)
\ArrowLine(150,89)(250,021)
%\ArrowLine(175,89)(250,021)
%\ArrowLine(200,89)(250,021)

\GCirc(50,021){5}{0}
\GCirc(100,021){5}{0}
\GCirc(150,021){5}{0}
\GCirc(200,021){5}{0}
\GCirc(250,021){5}{0}
\Text(320,000)[rb]{\Large output neurons}
\end{picture}
\end{center}

\caption{
Schematic layout of the Artificial Neural Network used for the primary
particle mass determination using the preprocessed muon data.
}
\label{anna1}
\end{figure}
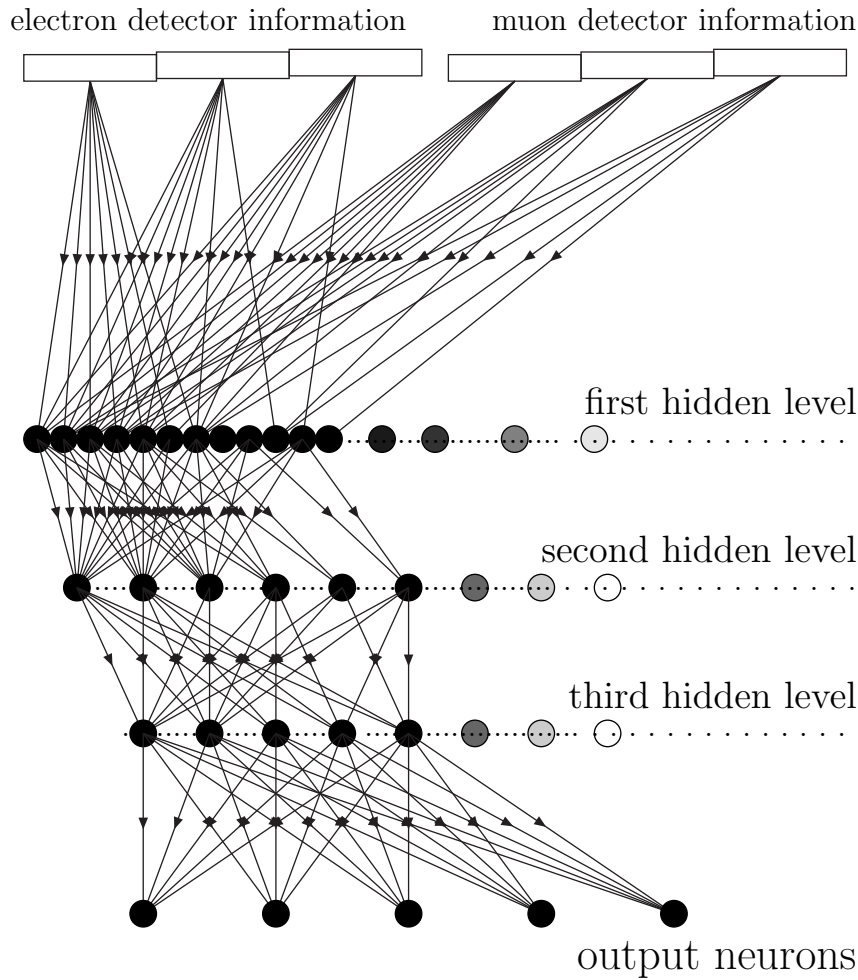

The structure of the network is shown in Fig.\ \ref{anna1}.

%%%%%%%%%%% ANN for A(mala)%%%%%%%%%%
%%%%%%%%%%%%%%%%%%%%%%%%%%%%%%%%%%%%%%%%%%

\begin{figure}
%\vspace{6cm}
\centerline{
\epsfxsize 130pt
\epsfbox{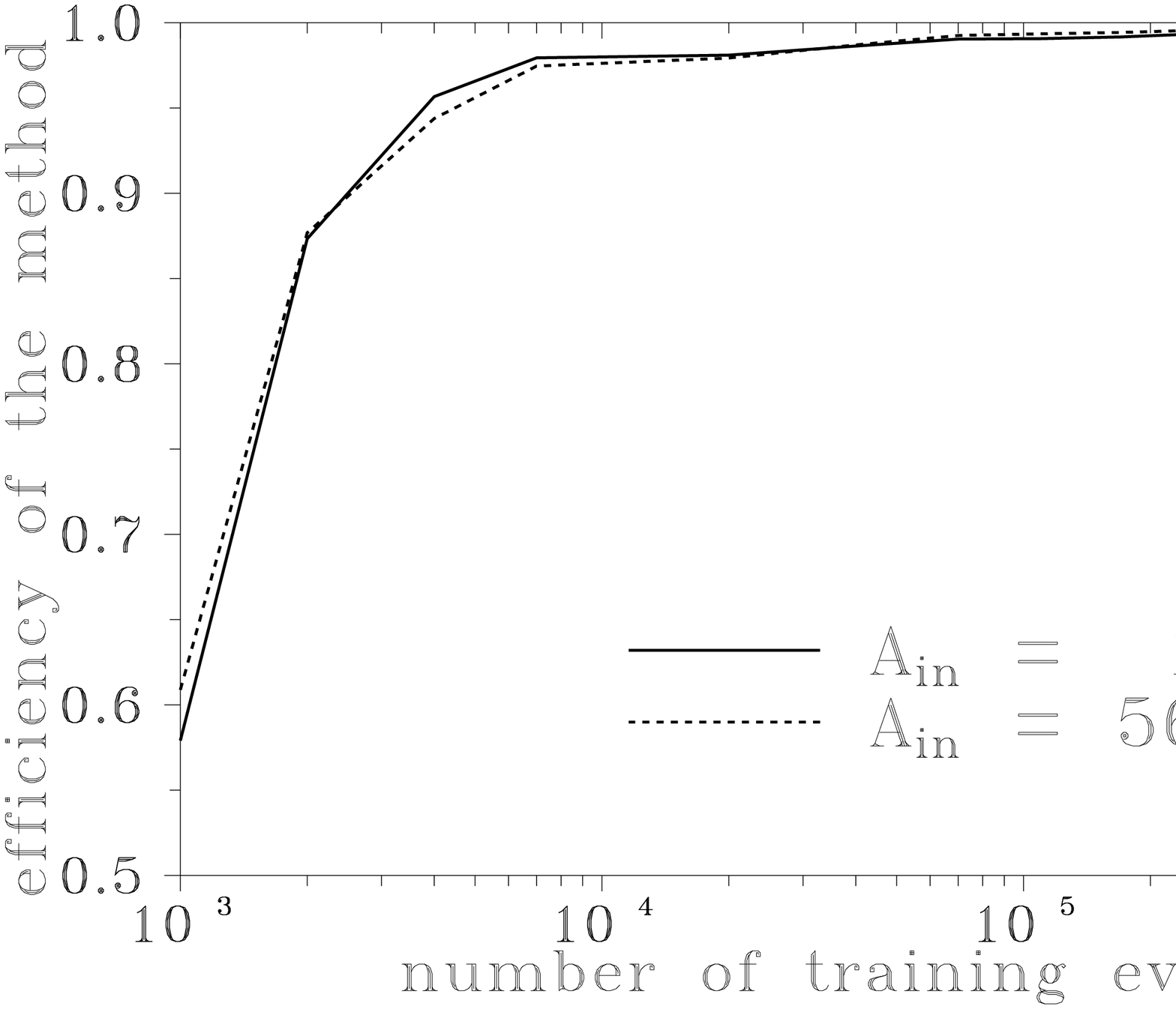}
\hspace{2cm}
%\vspace{-1.cm}
\epsfxsize 150pt
\epsfbox{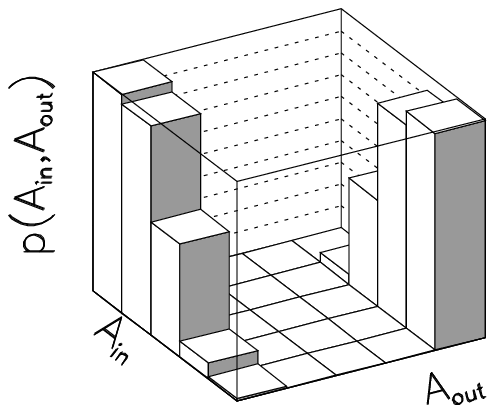}
}
\caption{
The results of separation of H and Fe by ANN using preprocessed muon data.
The efficiency of the ANN as a function of number of showers used for the
network training (left) and the final separation results (right).
}
\label{m1-56}
\end{figure}

First the behaviour of the network was tested for different pairs of five
possible masses in primary spectrum. This was done to see if the ANN is
able to distinguish between such as close events as produced
for example by H (A=1) and He (A=4). The physical fluctuations in the
shower development could disperse the information
about the primary mass. It was concluded that the efficiency
of the method in such a cases is rather questionable. However for outermost
masses in the primary mass spectrum: H and Fe the separation is almost
perfect as it is shown in Fig. \ref{m1-56}. The network efficiency is
defined as a conditional probability that the
ANN output is right for each particular primary particle mass.

\begin{figure}
\centerline{
\epsfxsize 130pt
\epsfbox{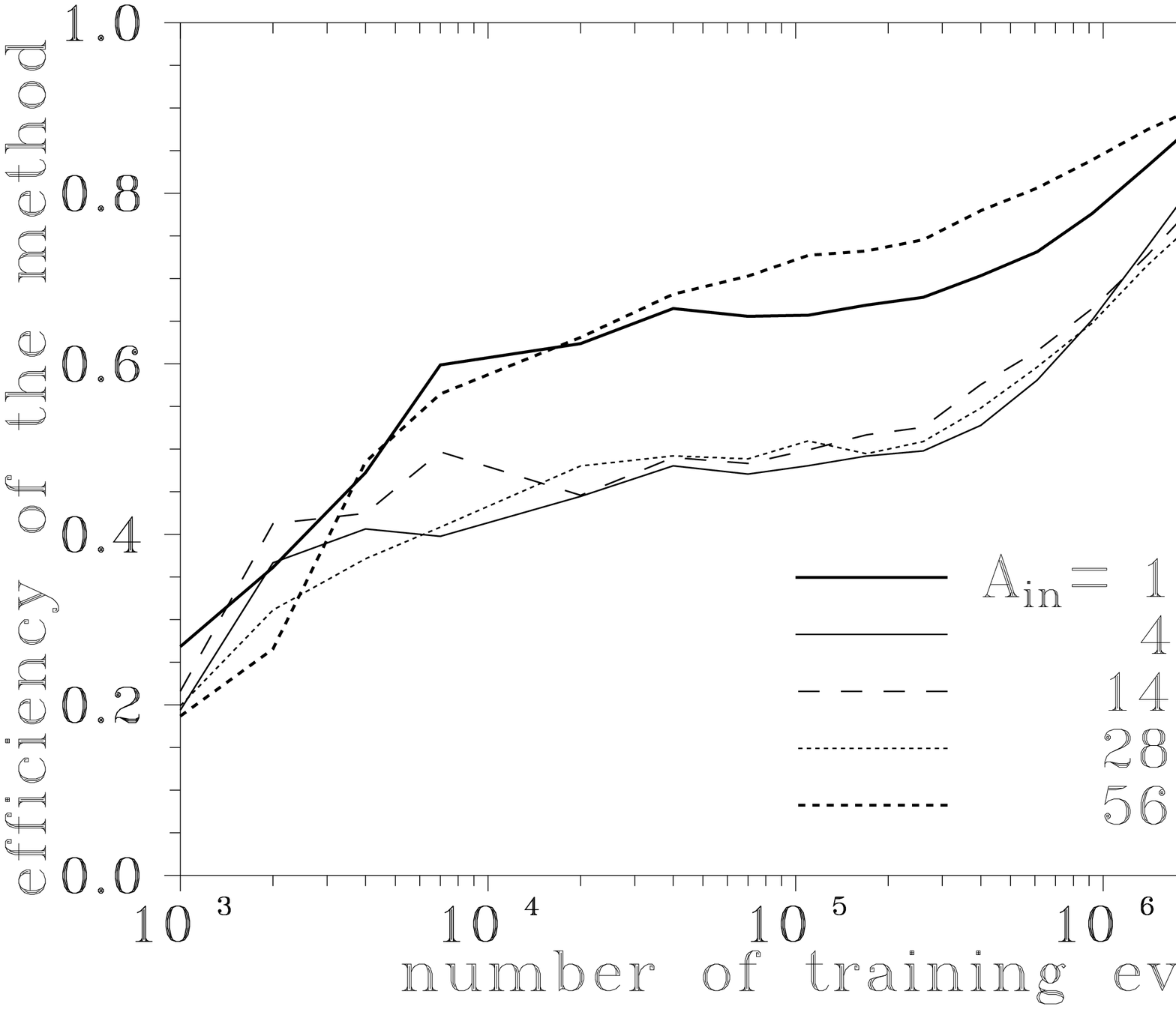}
\hspace{2cm}
%\vspace{-1.cm}
\epsfxsize 150pt
\epsfbox{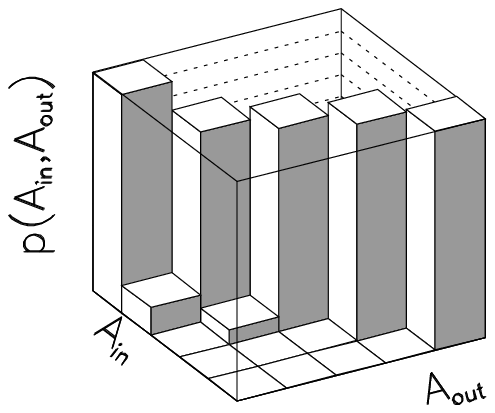}
}
%\vspace{6cm}
\caption{
The results of ANN method using preprocessed muon data for all five component
trained network.
The efficiency of the ANN as a function of number of showers used for the
network training (left) and the final separation results (right).
}
\label{mall}
\end{figure}

Difficulties with the close masses separation oblige to modify the
training procedure for the network training with the five component mass
spectrum. Requirement that the ANN answer should be exactly the one known
from the simulation {\em true} primary particle mass gives the training
process unsuccessful. Thus it has been replaced
by the broader presumed output with the maximum at the {\em true} value
but neighborhood output neuron signals were assumed to be higher
than the much distant ones. With such modification the ANN was trained with
all five components spectrum and the convergence was found. The results
are given in Fig. \ref{mall}.

The one more interesting possibility of the ANN architecture was examined.
There were used as an input signals raw detector outputs from all 192
muon detectors and as before the preprocessed electron data.
The statistical weight of an information about muons in each shower is reduced
due to the fact that the fraction of the muons registered by all detectors
in the KASCADE--like geometry experiment is about percent of all muons in the
shower. The results presented previously were obtained using the all muon
derived characteristic. The physically important question is if the
ANN method can be used directly with raw experimental data.

\begin{figure}
%\vspace{6cm}
\centerline{
\epsfxsize 130pt
\epsfbox{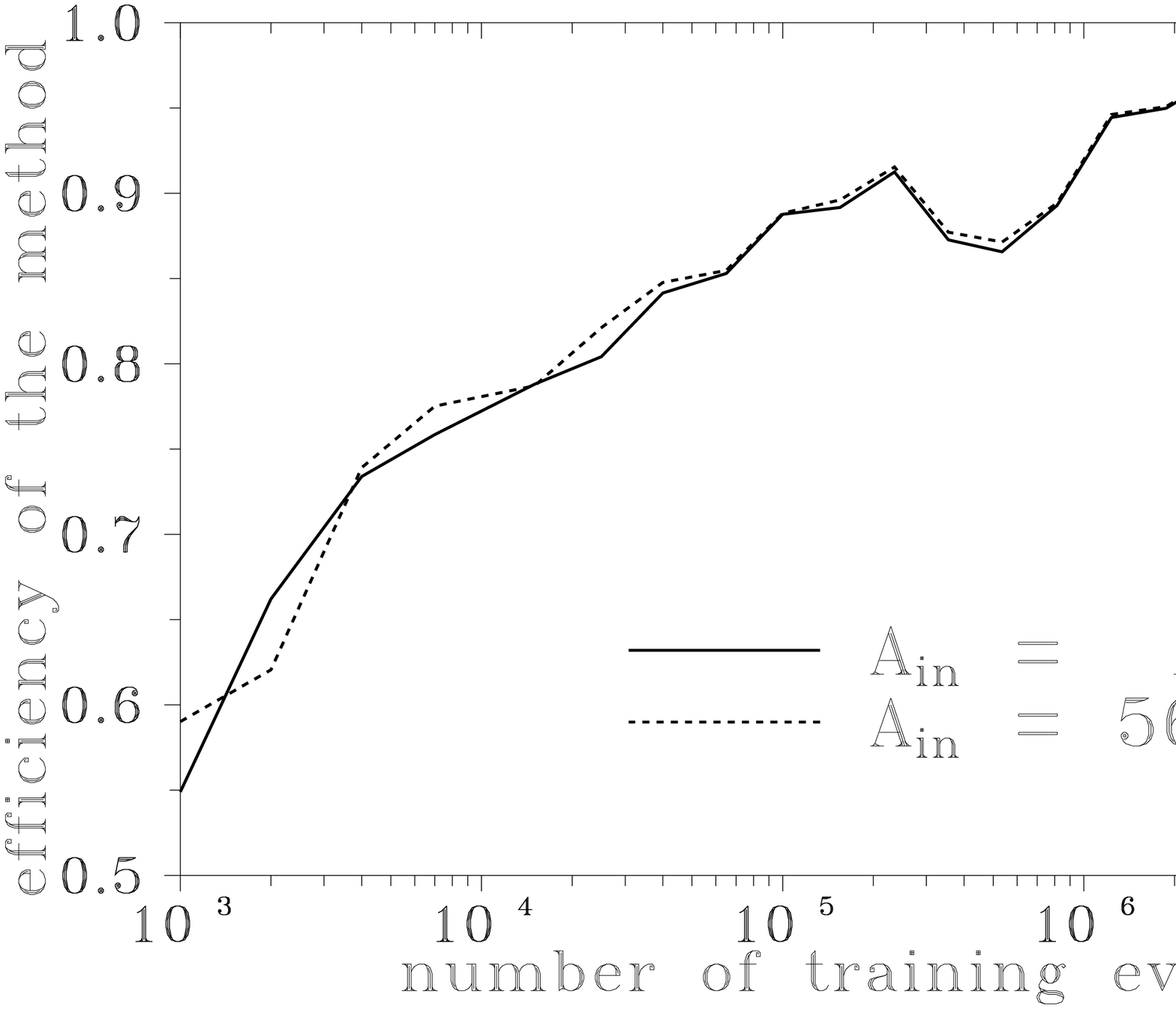}
\hspace{2cm}
%\vspace{-1.cm}
\epsfxsize 150pt
\epsfbox{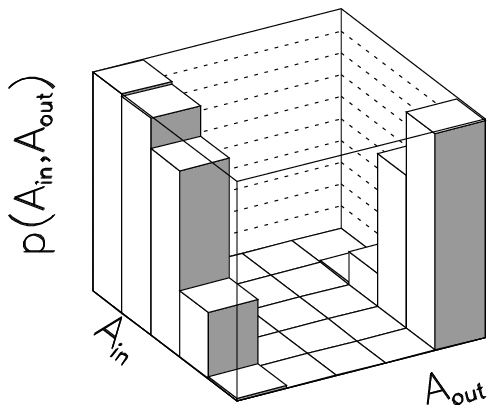}
}
\caption{
The results of separation of H and Fe by ANN using raw muon detector data.
The efficiency of the ANN as a function of number of showers used for the
network training (a) and the final separation results (b).
}
\label{d1-56}
\end{figure}

First again the H -- Fe separation possibility was tested.
The results are presented in Fig. \ref{d1-56}.

It is surprising that, in contrary to what one could expect, the final
efficiency achieved is not much worst than that obtained previously
(Fig. \ref{m1-56}). The only one what has changed is the training time.
The network needs much more simulated showers to reach the final resolution.
This is partially due to the increase of the number of neurons in the net.
%As it was expected the efficiency drops in comparison with the results in
%Fig. \ref{m1-56}, but is still surprisingly high.

\begin{figure}
%\vspace{6cm}
\centerline{
\epsfxsize 130pt
\epsfbox{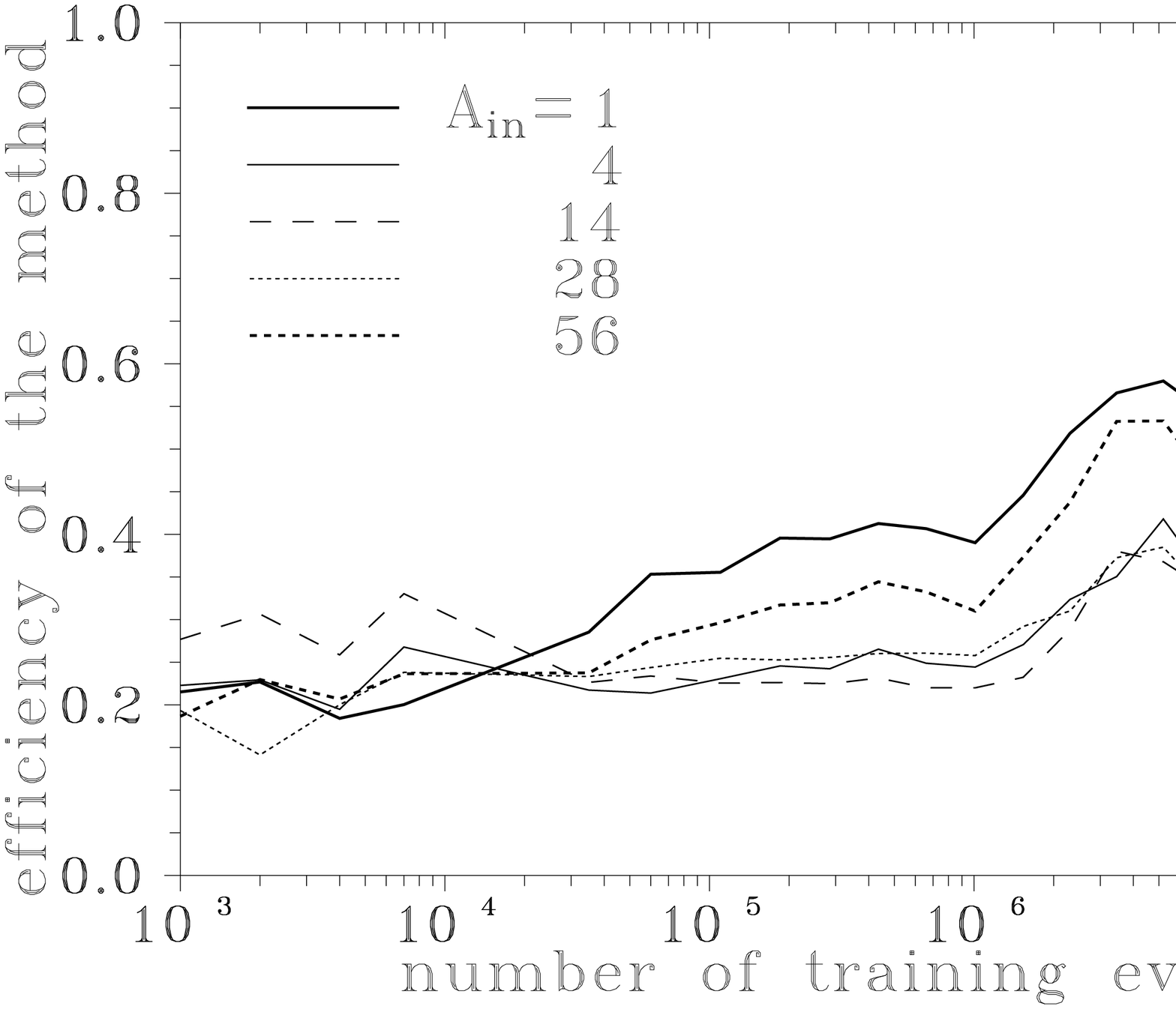}
\hspace{2cm}
%\vspace{-1.cm}
\epsfxsize 150pt
\epsfbox{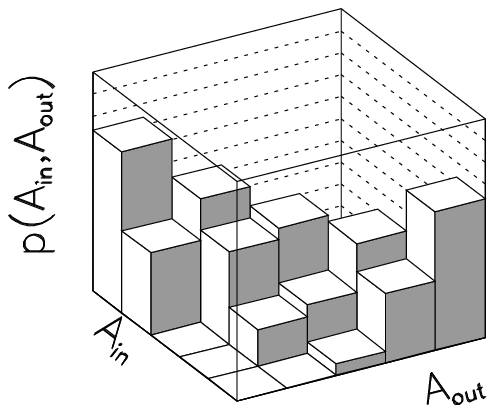}
}
\caption{
The results of
the primary mass separation by ANN using raw muon detector data.
The efficiency of the ANN as a function of number of showers used for the
network training (a) and the final separation results (b).
}
\label{dall}
\end{figure}

Next and final step of the present analysis is to see, if the raw muon
detector data from the KASCADE--like experiment allows one to distinguish
between different components in the whole primary mass spectrum.
The results are given in Fig.~\ref{dall}.

As it can be seen the efficiencies are worst than obtained previously
for the ''all muon data'' (Fig.~\ref{mall}). However it is clear that the
ANN is able to give some information about the primary particle mass.

The comparison of the efficiencies given in Figs. \ref{mall} and \ref{dall}
shows that the ANN approach is very effective and the
information collected by the detectors in KASCADE--like geometry experiment
is rich enough to distinguish between the showers initiated by primary protons
and iron nuclei only using the muons (and electrons) registered by the array
detectors.

\section{Summary}

The results presented in that paper show that the ANN method can be used
to find a total muon number in extensive air showers
as well as to classify showers due to primary particle mass.

The comparison of the total muon number obtained using ANN
with the standard minimization technics based method shows that
the ANN procedure works as good and sometimes even better (for the small
and very small showers) than the best minimization method.

The discrimination between different primary CR particle masses
looks very promising. The further careful study on improvement the
ANN sensitivity as well as on the accuracy of the EAS Monte--Carlo
generator is needed to create a powerful tool for data analysis
in cosmic ray experiments.

\end{document}